**Removing System Noise from Comparative Genomic Hybridization Data by Self-Self Analysis**

Yoon-ha Lee[1] (leey@cshl.edu)
Michael Ronemus[1] (ronemus@cshl.edu)
Jude Kendall[1] (kendall@cshl.edu)
B. Lakshmi[1,3] (lakshmi.muthuswamy@oicr.on.ca)
Anthony Leotta[1] (leotta@cshl.edu)
Dan Levy[1] (levy@cshl.edu)
Diane Esposito[1] (esposito@cshl.edu)
Vladimir Grubor[1,2] (vladimir.grubor@duke.edu)
Kenny Ye[4] (kye@aecom.yu.edu)
Michael Wigler[1,*] (wigler@cshl.edu)
Boris Yamrom[1] (yamrom@cshl.edu)

[1]Cold Spring Harbor Laboratory, 1 Bungtown Road, Cold Spring Harbor, NY 11724 USA
[2]Present Address: Institute for Genome Science & Policy, Duke University, Durham, NC 27708 USA
[3]Present Address: Ontario Institute for Cancer Research, Toronto, Ontario, Canada M5G 0A3
[4]Department of Epidemiology and Population Health, Albert Einstein College of Medicine, Bronx, NY 10461 USA

*To whom correspondence should be addressed.

**ABSTRACT**

**Background**

Genomic copy number variation (CNV) is a large source of variation between organisms, and its consequences include phenotypic differences and genetic disorders. CNVs are commonly detected by analysis of data created by hybridizing genomic DNA to microarrays of nucleic acid probes. System noise caused by operational variability and probe performance variability complicates the interpretation of these data.

**Results**

To minimize the distortion of genetic signal by system noise, we have explored the latter in an archive of hybridizations in which no genetic signal is expected. This archive is obtained by comparative genomic hybridization (CGH) of a sample in one channel to the same sample in the other channel, or 'self-self' data. We show that these self-self hybridizations trap a variety of system noise inherent in sample-reference (test) data. Through singular value decomposition (SVD) of self-self data, we are able to determine the principal components of system noise. Assuming simple linear models of noise generation, we present evidence that the linear correction of test data with self-self data—which we call system normalization—reduces local and long-range correlations as well as improves signal-to-noise metrics, yet does not introduce detectable spurious signal. Using this method, 90% of hybridizations displayed improved signal-to-noise ratios with an average increase of 7.0%, due mainly to a reduced median average deviation (MAD).

**Conclusions**

By applying system normalization to test data using an archive of self-self hybridizations with no genetic signal, we have been able to improve the performance of microarray CGH. In addition, we have found that principal component loadings correlate with specific probe variables including array coordinates, base composition, and proximity to the 5´ ends of genes. The correlation of the principal component loadings with the test data depends on operational variables, such as the temporal order of processing and the localization of individual samples within 96-well plates.

**BACKGROUND**

Genomic copy number variation (CNV) is a large source of variation between organisms [1-3]. The consequences of this variation include major phenotypic differences within members of the same species and highly penetrant genetic disorders [4-6]. CNVs are commonly detected by analysis of data created by hybridizing genomic DNA to microarrays of nucleic acid probes [4, 5, 7]. But extensive signal variation from hybridization to hybridization and from probe to probe complicates these analyses.

One common class of array platform is comparative genomic hybridization (CGH), in which two genomes—a sample and a reference—are simultaneously hybridized to the same array and hybridization intensity read as 'two-color' probe ratios formed from separate fluorescent channel intensities. One familiar form of system noise in this probe ratio data is evident as strong local trends when ratios are viewed in the genome order of the probes; another is long-range correlated probe biases, in which genetically unlinked probes have ratios that vary similarly in unrelated individuals across a study population [8-11]. Due to the law of independent segregation, there should not be any significant long-range correlation in the data; any long-range correlations present in the data should be due to system noise caused by various degrees of similarity between probes, DNA preparations and microarray treatments. The extent and types of such noise vary from hybridization to hybridization.

System noise is often best assessed when there is no true signal, hence we created and explored an archive of hybridizations in which no genetic signal is expected. This archive was obtained by CGH of reference in one channel to the same reference in the other channel. These experiments are known as 'self-self' hybridizations [12-14]. We show in this manuscript that self-self hybridizations trap a variety of system noise present also in sample-reference (test) data. The same local trends seen in some sample-reference hybridizations (Figure 1A) are also observed in some self-self hybridizations (Figure 1C).

Yet in some experiments, these trends are not in evidence (Figure 1B). In this work, we apply simple linear models of noise generation, and we present evidence that the linear correction of test data with self-self data—which we call system normalization—reduces trends and long-range correlations and improves signal-to-noise metrics, yet does not introduce detectable spurious signal in experiments that do not show local trends or long-range correlations. This is illustrated in Figure 1D–F. We use singular value decomposition (SVD) of the self-self data to determine the principal components of system noise [15]. The loadings of the principal components correlate with probe variables, such as discrete physical location of the probes on the microarray surface, base composition and proximity to genes [8, 10]. The correlations of the principal components with the test data also depend on operational variables, such as the sample queue index (which reflects batch variability in processing). Operational variables may also act within a batch, such as those corresponding to the locations of DNA samples within the 96-well plates in which they are stored and shipped.

## METHODS

**Origin of test and self-self ratio vectors**

Our data set consists of a group of 3252 test (sample-reference) hybridizations and a group of 132 self-self hybridizations. The latter group was comprised of 83 self-self hybridizations of our standard human male reference genome and 49 self-self hybridizations of other sample genomes, chosen at random. All test hybridizations were performed with the same male reference DNA and the same choice of dye labels: Cy3 for the sample and Cy5 for the reference. The self-self group consists of hybridizations with various batches of reference DNA or sample in both channels. The self-self hybridizations were randomly interspersed among a larger set of CGH experiments performed over a period of ~1 year. Blood samples were collected at a variety of centers throughout the United States. Sample and reference DNAs were prepared either from whole blood or from EBV-immortalized B-cells at the Rutgers University Cell and DNA Repository (RUCDR). DNAs were prepared robotically, then distributed and stored in 96-well plates. We track the reference batch number and the sample queue indices (multi-well plate, column and row). All hybridizations were performed by NimbleGen in their Icelandic facility. DNAs were labeled by random priming incorporating a fluorescent cytosine nucleotide derivative. The platform was a NimbleGen HD2 CGH microarray with 2.1 million probes, the positions of which were randomized across the array surface. Composition and locations of probes on the array were kept fixed throughout the period of data collection.

We do not perform background subtraction. Rather, we employ other steps in data processing that are commonly used in the field, namely local and Lowess normalization of probe intensities [14, 16]. We will refer to the natural logarithm of ratios of such normalized probe intensities—when placed in genome order—as the local and Lowess normalized 'ratio vectors,' omitting various modifiers when the meaning is clear. When we remove the data from the X and Y chromosomes, we refer to the remaining data as autosomal ratio vectors.

We segment ratio vectors into distinct regions of constant copy number using Kolmogorov-Smirnov (KS) statistics to determine if the segmentation passes the threshold of significance [17]. The results we discuss are essentially unchanged if we use other segmentation procedures such as circular binary segmentation [11].

**Mathematical ideas behind system normalization with the self-self archive**

The data $Y_i^k$ represent the local and Lowess normalized log ratios. Probe index $i$ ranges from 1 to $N$ ($N = 2161679$) and hybridization index $k$ ranges from 1 to $M + L$, where $M = 3247$ is the number of test hybridizations and $L = 132$ is the number of self-self hybridizations. Omitting the index $i$ and assuming a linear additive noise model, we write

$$Y^k = G^k + S^k + \varepsilon^k, \qquad (1)$$

where $G^k$, $S^k$, and $\varepsilon^k$ are unobserved vectors in the $N$ dimensional linear vector space $W$. $G^k$ is the genetic signal vector representing copy number differences between the sample and the reference, a piecewise constant function of the probe index $i$ for each hybridization $k$. $S^k$ is the major system noise vector; and $\varepsilon^k$ is residual noise.

Our goals are two-fold: to improve the estimate of $G^k$ by correcting $Y^k$ as best we can for system variability; and to study the structure of the system noise for insights into its origins. We start by exploring system variability using singular value decomposition in self-self hybridizations, where $G^k$ is zero. We then proceed by correcting this variation in test hybridizations. For these hybridizations the singular value decomposition of the $N$ by $L$ matrix, $A$, composed from columns $Y^k$, is

$$\qquad (2)$$

where $U$ is a $N$ by $L$ matrix with orthonormal columns, $D$ is an $L$ by $L$ diagonal matrix with non-negative singular values on the diagonal; and $V$ is an $L$ by $L$ matrix with orthonormal columns, and $V^T$ is its transpose. Singular values decrease sharply, which indicates that most of the variation in self-self hybridizations is concentrated in a lower dimensional subspace $U'$ spanned by the first few columns (major principal components) of matrix $U$. To correct $Y^k$ for system noise, we subtract from $Y^k$ its orthogonal projection $Y_{U'}^k$ to $U'$. We show that this greatly reduces local trends, long-range correlations, and improves signal-to-noise in corrected self-self hybridizations.

We next assume that the components of system noise captured by the self-self hybridizations (and described by the principal components) are also shared in test hybridizations, and correct the system variability from the latter by subtracting from them their projection onto the subspace $U'$. As a practical matter, to compute the coefficients $\alpha$ of the orthogonal projection to $U'$ in terms of the principal components, we use only the probes from the autosomal region of the genome. This circumvents the distortion in the projection that would be caused by large areas of genome with known difference in copy number between the sample and the reference when the sample is from a female (the unavoidable consequence of using a male reference genome). After computing the coefficients $\alpha$ on this truncated basis, we apply $\alpha$ to the full-length basis to compute the projection for the probes on the X and Y chromosomes. As a side benefit, we can evaluate the performance of system normalization upon the data for the sex chromosomes, which data is not involved in computing the correction.

## RESULTS

**System normalization using self-self vectors**

Self-self ratio vectors contain system noise, most readily seen as long-range correlations between probe ratios. To view the extent of these correlations, we sampled 2000 probe ratios at random, avoiding probes with multiple mappings to the genome, and then computed their Pearson correlations across all self-self data. A histogram of these values is shown in Figure 2A (blue curve). To test system normalization on these vectors, we used singular value decomposition [18] to break down the self-self vector set into the principal components, a set of orthogonal vectors from which we pick a basis for the subspace $U'$ (as described in the Methods). To determine the number of major principal components—those with the largest singular values—we compared the singular values from actual self-self ratio vectors to those from the within-row-permuted $N$ by $L$ matrix, $A$, of self-self vectors. The ratios within each row were independently permuted, thus obliterating the correlations between probe ratios arising from system variables but maintaining the mean and standard deviation for each probe ratio within the self-self archive. The comparison suggested taking the first 14 principal components to generate the subspace $U'$ (see "singular values," Table 1 and Figure 3). Another method, the Scree plot, suggested the same number of major principal components [18]. Correcting the self-self hybridizations by their projection to $U'$ greatly reduced long-range correlations in the data (Figure 2A, red curve). After noise reduction, the long-range correlations are comparable to what is seen following within-row-permutation of matrix $A$ (Figure 2A, green curve).

We next applied this method to test vectors, using the residual after removing their projection onto $U'$ as described in the previous section. As a first test that this method works, we again sampled the long-range correlations of probe ratios in the test vectors using the same method employed for self-self vectors (see Figure 2B, same color codes as above). There are also substantial decreases in the extent of long-range correlations in test data (compare blue to red). This method of system normalization, using the fourteen major principal components, will be designated MPC.

**Assessing performance of system normalization**

For test hybridizations, measures of signal and noise are readily available. When the sample is female (given the reference is male) the median of the ratio on the X chromosome, excluding the pseudo-autosomal regions, is an obvious measure of signal. We use the X-specific signal when measuring signal to noise, in part because the correction of the X for system noise is based on projections from autosomal data. We use the median absolute deviation (MAD) of probes from the X chromosome, excluding the pseudo-autosomal regions, as the simplest measure of noise. We call the ratio of X-specific median to X-specific MAD the 'signal-to-noise,' and it serves as one of several guides as to whether a normalization protocol improves performance. When system normalization was applied, median signal decreased in 63% of all test hybridizations. But MAD was reduced to a greater degree, and 95% of all test hybridizations showed decreases (which averaged 6.8%). In all, 90% of hybridizations displayed improved signal-to-noise ratios, with an average increase of 7.0%. For 20% of hybridizations, the level of this improvement exceeded 10%. This is also shown qualitatively as a histogram of signal-to-noise for our local and Lowess ratio vectors and their corrections (Figure 4A). For samples of any gender, other more involved measures of signal on the autosomes are available using regions of common polymorphism. The autosomal and the X-specific signal measures are in excellent agreement (data not shown).

The MAD gives an overall measure of noise, but there are multiple types of noise that display different patterns, such as local trends, excessive segmentation and label-biased ratio vectors. We use three measures for these three noise patterns. The simplest measure for local trends is autocorrelation, which is the Pearson correlation of the ratio vector with itself shifted by one index. A histogram of the autocorrelation for ratio vectors is shown in Figure 4B. A histogram of the number of segments following KS segmentation is shown in Figure 4C, where excessive segmentation is seen as a thick rightward tail. A measure of label bias is the ratio of the number of segments scored as amplifications to the number scored as deletions. A histogram of this measure is shown in Figure 4D.

We examined these measures of noise after MPC correction of test vectors. All the measures of system noise in test hybridizations diminish (Figure 4A–D).

One way to gauge the impact of system normalization is by examining the frequency with which certain regions of the genome are found segmented. The most marked change is seen in low-amplitude events (Figure 5A). For each probe, we calculate how often it is observed contributing to a segment with a median ratio above a given threshold. We plot segmentation counts at each probe corresponding to the autosomes, using a ratio threshold of log(1.1), as determined from the set of 3252 test hybridizations, plotted as 'before' (X-axis) vs. 'after' (Y-axis) MPC normalization. The frequency of a large set of low amplitude segments detected before system normalization is drastically reduced after normalization (graphical region marked "A"). We think of this change as arising from reduction of genomically clustered system noise (see later sections) that produce low amplitude segments. The frequency of a few common copy number polymorphisms decrease modestly (graphical region marked "B"), and the probes from these regions often overlap with regions in our reference genome where we strongly suspect it has copy number zero (data not shown). We do not see entirely new regions of segmentation that become common only after system normalization, as would likely be the case if artifacts were being introduced. On the other hand, the frequency of detection of many more common events actually increases modestly, which we think happens because of improved signal-to-noise in some of the noisier hybridizations (graphical region marked "C").

Another way to gauge the effectiveness of normalization is by examining the clarity of underlying copy number states. For any region of common copy number polymorphism, the variation should be observed as discrete states in the human population corresponding to actual quantal increments of copy numbers. In fact, multiple distinct states were readily observed in several commonly polymorphic sites only after system normalization. An example of one such region, chosen from a subset of CNPs of >10% frequency in the sampled population, is shown in Figure 6. Before system normalization, four peaks representing distinct copy number states are apparent (lower left and middle panels). After system normalization, at least six discrete copy-number states could be distinguished, and the fourth (rightmost) state is resolved into three states (lower right panel). Overall, we found that system normalization greatly improved the clarity of state.

**Association of probe variables with principal components**

Examining the properties of the principal components chosen for MPC, in particular their associations with known system variables, reveals a surprising richness and structure to system noise.

We first examined the association of the probe loadings of each principal component with three probe properties: the location of probes on the microarray surface, the location of the probes in the genome, and the nucleotide composition of the probes. Nine components, the fifth through eighth and the tenth through fourteenth, exhibit strong spatial clustering of probes of extreme loadings (highest 1.5% and lowest 1.5% of values) on the array surface (Figure 7 and Table 1). The spatial cluster patterns reflect that the arrays are printed in three blocks, and each block is processed in separate hybridization chambers.

The probe sets with extreme loadings show statistically significant compositional bias compared to random sets of probes. The total nucleotide difference (defined as in Table 2) of 1000 sets of randomly selected probes ranges from 0 to 0.059, but those of probes with extreme loadings of our 14 principal components range from 0.0238 to 0.3138. The first, second and fifth components stand out in this regard. For the first (and fifth) component, the base composition of probes with extreme high loadings is strongly enriched for both C and G and depleted for A and T, and the reverse is true for the probes with extreme low loadings. The physical basis for the first component may thus lie in the strength of duplex binding. Because the binding energy of duplex formation is strongly dependent on the proportion of C:G pairing [8], hybridization efficiency is sensitive to temperature and salt, and perhaps even shearing forces. For the second component, probes with extreme high loadings are strongly depleted in A and enriched for T relative to the probes with extreme low loadings, with C and G being unaffected. Although a swap of A for T would not alter duplex stability, it might alter the chemical properties of probes. We can speculate, for example, that many of the extreme probes of the second component, enriched for A and depleted for T, differentially quench the fluorescence in the Cy3 and Cy5 channels, with the reverse true for those enriched for T and depleted in A.

Four components, the first, third, fourth and ninth, show strong autocorrelation (see Table 1), with the first (0.35) and ninth (0.33) the strongest, compared to other components such as the second (0.05). Autocorrelation leads to spurious segmentation. Hence, these components are of greatest interest to us, as they can lead to false positive segments. As is well known, the C+G base composition is not randomly distributed in the genome [19-21]. The autocorrelation of the first component probably reflects the G/C bias of the genome; because the G/C is clustered, so is the noise captured by this component (Figure 8A).

The explanation of the autocorrelation for the ninth component was unexpected, and is likely more biological than physicochemical. This component has strong autocorrelation, but not an exceptional compositional bias. Unlike the other components, it has high skewness and excess kurtosis (Table 1). The distribution of the loadings has a long one-sided tail. The probes from this tail have a remarkable distribution in the genome: they tend to cluster near the transcriptional start sites of genes (Figure 8B–D) that also contain CpG islands [22]. Here we define a probe cluster as a maximally contiguous set of at least three extreme probes from the top 1.5% of loading values, and we define the probe cluster interval as that spanning the first and last probes. With these definitions, there are 3415 cluster intervals, 57% overlap the 5´ end of a gene, 68% overlap CpG islands, and 54% overlap both. Such level of overlap is highly unexpected based on our simulations. We randomly created 3415 new probe clusters from our probe set and recomputed the percentage of overlap with the 5´ ends of genes. In 100 simulations, the overlap ranged from 5 to 7%. The observed overlap, 57%, lies far outside this range, and its p-value is certainly far below $10^{-2}$. Nor does the ninth component follow the G/C composition of the genome (Figure 8A). None of the extreme probes of the other components form as many probe clusters nor probe clusters with this pattern (Tables 1 and 3).

**Association of operational variables with principal components**

The production of the hybridization vector depends upon several operational variables: the cell source; preparation and measurement of DNA; the synthesis of microarrays; the hybridization and wash conditions; and the settings and conditions of microarray scanning. For our pipeline, all of this occurs at multiple different centers at which: 1) blood cells are collected; 2) DNA is made from those cells; 3) arrays are fabricated; and 4) hybridizations are performed and arrays scanned. A single variable can be used that captures much of this operational complexity, namely the 'queue index.' This index contains information about the order of processing, and the position of samples within multi-well plates. For samples delivered in 96 (8 by 12) well plates, the queue index is defined as the sum of the plate (or batch) number (in order received, processed and shipped, starting from zero to 32) times 96, plus the row number (starting from zero to 7) times 8, plus the column number (from 1 to 12). For each local and Lowess normalized test ratio vector, we computed the Pearson correlation coefficients with each principal component, and plotted these correlation coefficients as a function of the queue index of the samples. We also computed the correlation coefficients following MPC normalization. All fourteen plots are shown in Figure 9. For all components except the second, the correlation with local normalized ratio vectors range from positive to negative. The second component is unusual, as the sign of its correlation with the local normalized test ratio vectors remains unchanged throughout the entire period of data gathering. We think this component corresponds to what is usually called 'color bias', and is normally handled by repeating the hybridization with dye swap. For all components except the ninth, variation in correlation between the components and queue index was a rough function of the batch.

The correlation to the ninth component shows a different pattern. The correlation is not dependent on batch, but rather its variation has a periodicity of 12 with respect to the queue index (Figure 10A). A periodicity of eight is also apparent when the index is computed differently. The origin of this periodicity becomes evident by considering the placement of samples in the 8 x 12 micro-well format. To see this most clearly, correlation coefficients in each 96-well plate were normalized to a mean of zero and standard deviation of one, and then the correlation coefficients from wells with identical row and column numbers are averaged, yielding a heat map in which each well coordinate is a single block (Figure 10B). The extent of noise in this ninth component is clearly a function of well coordinate, in which the distance from the long and short edges of the plate are the critical variables. No other component displays this pattern.

**Batched correction for ninth component**

For all components, MPC normalization removes correlation with the principal components—as expected. However, we were especially concerned about the correction of the probes with extreme loadings from the first through fourth and the ninth components, as these components are major and/or are strongly autocorrelated. Thus they may contribute most to spurious segmentation. We examined correlation of the MPC normalized data with each component only at those extreme probes. The extreme probes are corrected well for the first four components (not shown), but not for the ninth component (Figure 10A).

The ninth component strongly affects a very small set of probes: a sufficient number to be detectable as a principal component, but an insufficient number in any given hybridization vector to force correction against the contravening introduction of white noise caused by the correction. This conclusion motivated us to try a non-linear treatment of the data. We ranked all probes by their loadings in the ninth component and grouped probes in batches of 50,000 by their rank, thus partitioning the probe set. The probes with high loadings in the ninth component—and therefore highly sensitive to that system noise component—are thus heavily represented in the first batch of probes. We applied MPC normalization to each batch of probes separately, treating them as 'mini-genomes' with their autosomal part equal to the intersection with autosome probes of the whole genome and their X, Y part equal to the intersection with probes on X and Y chromosomes. Normalized batches are assembled afterwards into the whole genome.

The results of this method are virtually indistinguishable from our other methods, except that the extreme probes from the ninth component are now better corrected (Figure 10A). We call this method 'batched principal component' normalization (BPC). For the vast majority of the probes, BPC is not distinguishable from MPC (Figure 5B). There is a small set of probes that are less frequently segmented (see circled set in Figure 5B). The correction for these component 9-sensitive probes is improved, but still is far from complete.

## DISCUSSION

We have been engaged in genetic studies of children affected with disorders (autism, congenital heart disease and pediatric cancer) born to otherwise healthy parents. We search these children for genomic copy number variants not seen in either parent (i.e., 'Mendel violators') because new variation seen in the child provides strong clues to the genetic origins of the disorders. Such *de novo* events are truly rare, so it has been critical for us to minimize false discovery rates. CGH often contains noise of many types that interfere with interpretation, so we have been highly motivated to reduce systematic noise as well as to understand its origins, and have chosen a method which does this well.

We detail methods to correct 'correlated' noise, which is easily recognized by two distinct signatures. The first signature is a pattern of local (in genome order) trends in ratio data within single hybridizations that often leads to spurious segmentation. The second signature consists of long-range pair-wise probe correlations between different experiments. Not all hybridizations show strong noise signatures and the signatures vary so it is not possible to correct such noise simply by mean adjustment [23]. On the other hand, taking the residual after projection to the space of the major principal components corrects data when the noise is abundant, and barely alters signal when such noise is not present. The cost of adding a few self-self hybridizations is minimal, less than 3% of the total hybridizations, and we have been able to omit the color reversal, which aside from giving another measurement of the same sample, probably corrects mainly for the second component.

Computing residuals to the principal components derived from the test data itself is problematic, because those principal components also contain genetic signal, namely the copy number differences between the genomes of the subject and reference genome. When PCA is based on sample-reference data, strong traces of common polymorphic genetic signals are found in all the major principal components, and therefore corrupt the corrections. To solve this, we hypothesized that major system noise is also present in self-self hybridizations. In self-self hybridizations, we expect no genetic signal, and so principal component analysis should reflect only system noise. We designed our data collection with self-self hybridizations liberally inserted into the production pipeline. We observed that much of the system noise that afflicts sample-reference hybridizations is also found in the self-self hybridizations, and therefore we could use the latter to correct for noise in the former.

As strong validation of our approach, the properties of the major principal components reflect known system and operational variables. For example, the extreme probes in several components reflect the layout of probes on the array, consistent with the expectation that some variation arises from fabrication and/or physical processing of the arrays. Also, extreme probes from the first two components have striking biases in their base compositions. The extreme probes of the first component are biased by C+G content. Because the efficiency of hybridization varies with C+G content of the probes, the first component may reflect imprecisely controlled hybridization and washing conditions. This component is also responsible for major trends in the data, as expected from the presence of C+G rich isochores distributed throughout the genome [19, 20, 24]. The probes of the second component have a bias in A at one extreme and in T at the other. The second component is the most invariant of all the components with respect to the operational variable of time, and hence it may reflect a physiochemical interaction of the probe with the fluors.

The most unique principal component is the ninth. The probes with extreme loadings of this component often map to intervals containing both the 5′ ends of genes and associated CpG islands. Yet these probes are not themselves especially rich in C/G. The origin of this variation is a mystery. The magnitude of this noise component is dependent on the coordinates of the sample in its 96-well plate. But samples are not initially prepared in 96-well order: after genomic DNA preparation, samples are assembled into 96-well order. Samples are shipped and subsequently processed retaining that order. Perhaps a feature of the chromatin structure surrounding certain regions leaves a footprint when DNA is prepared, and this footprint is seen only when it interacts subtly with subsequent DNA handling [25]. One possibility is that the footprint, perhaps DNA strand breaks, depurination, or residual chromatin complexes, creates a signal in combination with an unknown operational variable, such as thawing.

We show that test hybridization vectors can be partially corrected after computing their orthogonal projections onto the subspace of the major principal components. Although most hybridizations are of good standard quality and not in need of correction, the measures of performance as a whole improve upon correction. Overall signal-to-noise (measured on the X-chromosomes of female samples) improves, even though probe ratios from the X-chromosomes are not used to compute the correction. Pearson correlation coefficients between distant probe ratio pairs decrease as well, and segmentation improves by a number of measures.

Nevertheless, correction is not complete. There are still hybridizations that show excess segmentation, as well as truly awful signal-to-noise ratios. These hybridizations represent outright failures, and nothing can repair them. More troubling is the noise from the ninth component. Hybridization can have reasonable signal-to-noise and yet have distorted ratios in certain chromosomal regions leading to spurious segmentation. Until we realized this, we were puzzled by a set of apparent small copy number events in leukemias that we could not validate using other methods of copy number measurement. Eventually, we realized that these segments were all derived from the extreme probes of the ninth component. We can improve the correction by partitioning probes according to their loadings in the ninth component, and performing principal component correction on each partition separately. By concentrating probes that are noisy with respect to one component we can correct them better. But the correction is still not totally satisfactory, and that may be in part because there is a variable biological factor at play of great complexity.

The use of self-self hybridizations is central in this process, because such data are not contaminated by true signal. Although our method is based on the NimbleGen HD2 microarray platform, we expect similar methods would work on other microarray-based platforms or even on DNA sequencing-based methods for estimating copy number. We have not explored other mathematical methods for component analysis, and certainly principal component analysis is neither the only method nor necessarily the best. Other possibilities include independent component analysis [26, 27], sparse component analysis [28], partial least squares [29], and perhaps non-linear methods [30]; these alternate methods remain for further exploration.

**CONCLUSION**

We show that self-self hybridization captures much of the noise in micro-array genome copy number data (Figure 2). Data from self-self hybridizations can be configured as purely additive noise, and its principal components can be used to fit sample-reference data. The residual corrected data has substantially less noise, as judged by a variety of means. Using self-self is required in our method, as principal components derived from sample-reference data distorts true signal, severely interferes with segmentation at regions of common polymorphism, and erodes modeling of data as discrete copy numbers. The cost of a few self-self hybridizations is minor, compared to overall costs, less than 5% of the total number of sample-reference hybridizations, and probably obviates the need for color reversal, so is a large net saving. The other assets of our approach to noise reduction are many. We not only reduce false positives (Figure 4, Figure 5A), but also improve somewhat the detection of some common events (Figure 5C) and improve our ability to call discrete genetic states clearly (Figure 6). The observation of how the principal components depend on system variables can also be used to discover physical and temporal sources of noise in a data pipeline (see Figures 5, 9 and 10), and to reveal noise sources arising from probe composition (Table 1) and even the functional units of the genome (Figure 8). Because the principal components depend on probe, genomic and operational variables, our particular components are not applicable to another pipeline, but the general approach should be applicable whenever a portion of the output can be set aside to capture system noise.

**AUTHORS' CONTRIBUTIONS**

Y-HL and BY helped to conceive the study and carried out the main informatics and statistical analyses and helped to draft the manuscript. BL helped to conceive the study and carried out the initial data analysis. VG and DE contributed biological observations and interpretation. MR contributed biological observations and interpretation, and helped to draft the manuscript. AL and JK developed algorithms and assisted in the data analysis. DL assisted in the data analysis. KY helped to develop statistical methods. MW helped to conceive and coordinate the study and draft the manuscript. All authors read and approved the final manuscript.

## ACKNOWLEDGEMENTS

We thank the Simons Foundation Autism Research Initiative (SFARI) for their financial support of this work, and the Rutgers University Cell and DNA Repository and Roche NimbleGen, Inc. for their technical assistance. We would also like to thank David Donoho for helpful discussions.

## DATA ACCESS

Raw and processed data files corresponding to all hybridizations in this study are available in the Gene Expression Omnibus (GEO) at NCBI (http://www.ncbi.nlm.nih.gov/geo/). The GEO accession is GSE23682.

**FIGURES**

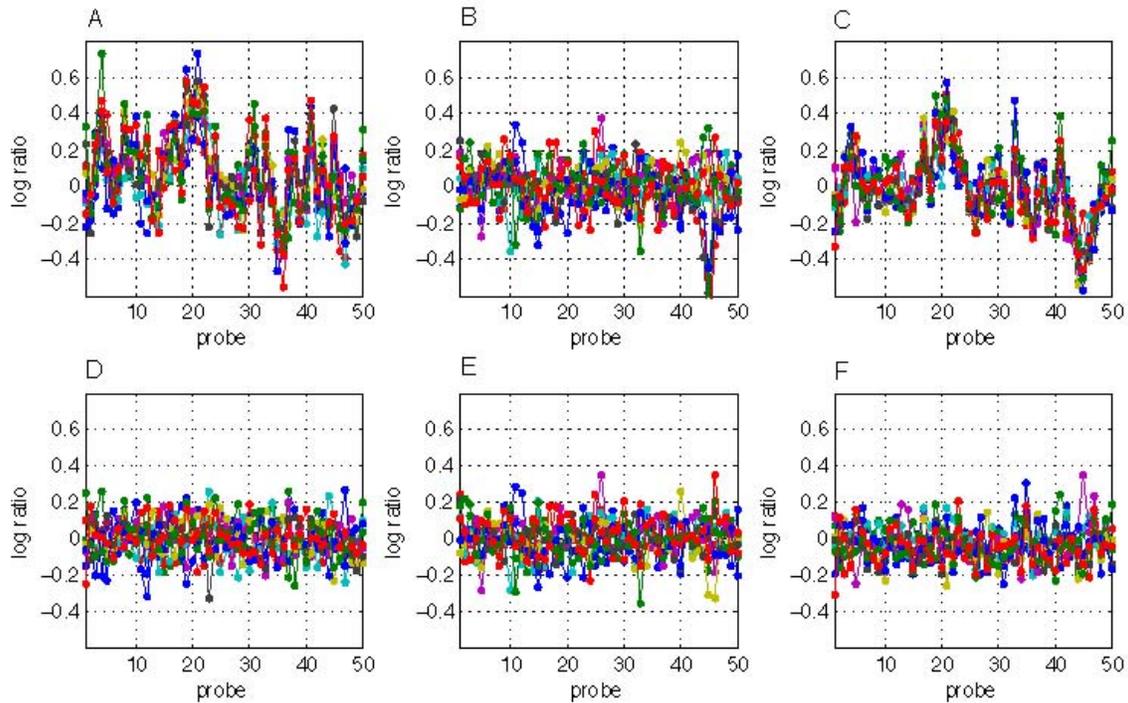

**Figure 1.** Treatment of system noise. To illustrate the problem and its solution, we selected one region of 50 contiguous probes without genetic signal from sample-reference and self-self hybridizations. (A) Ten sample-reference hybridizations in which visible trends are present; (B) the same region from ten more typical sample-reference hybridizations which do not have the trendiness artifact; and (C) this region in ten self-self hybridizations which do. Lower panels (D–F) show the corresponding experiments and region after system normalization by MPC (see text for further detail). All graphs display log ratios.

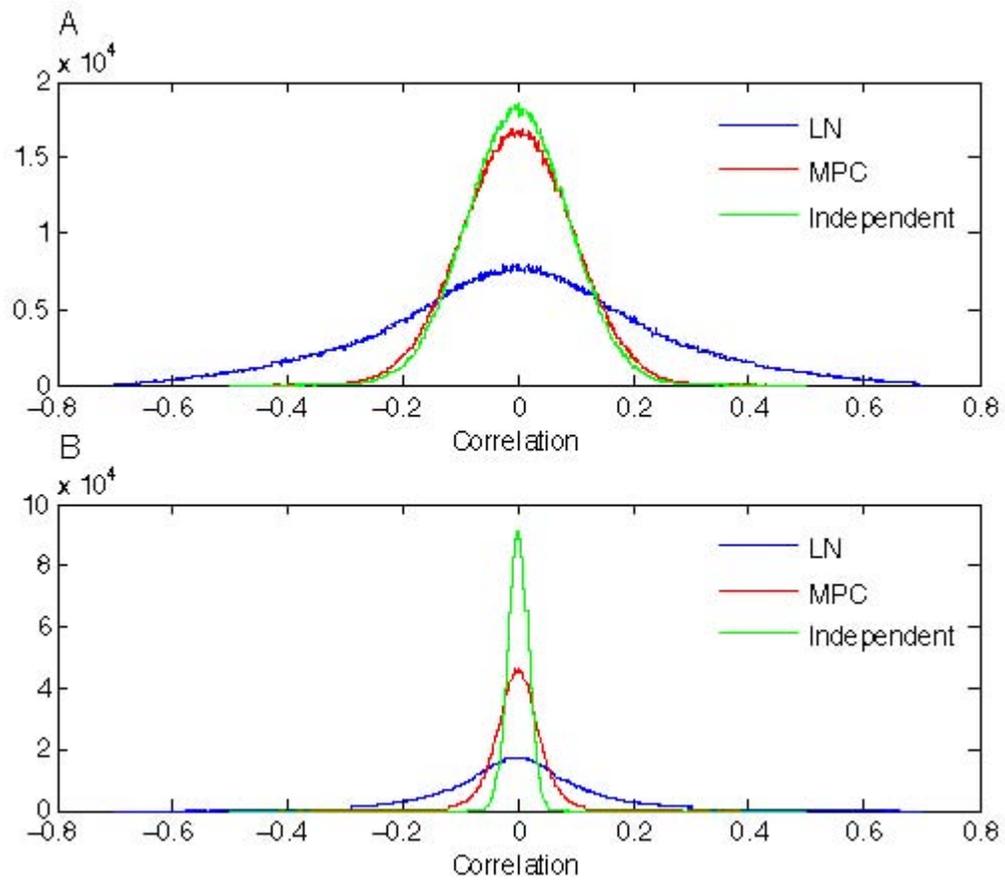

**Figure 2.** All pair-wise correlations of 2000 random probes. Histograms of correlations are shown in 132 self-self (A) and 3252 test-reference (B) hybridizations. LN are local and Lowess normalized values, MPC are major principal component corrected values, and "independent" refers to within-row-permuted values.

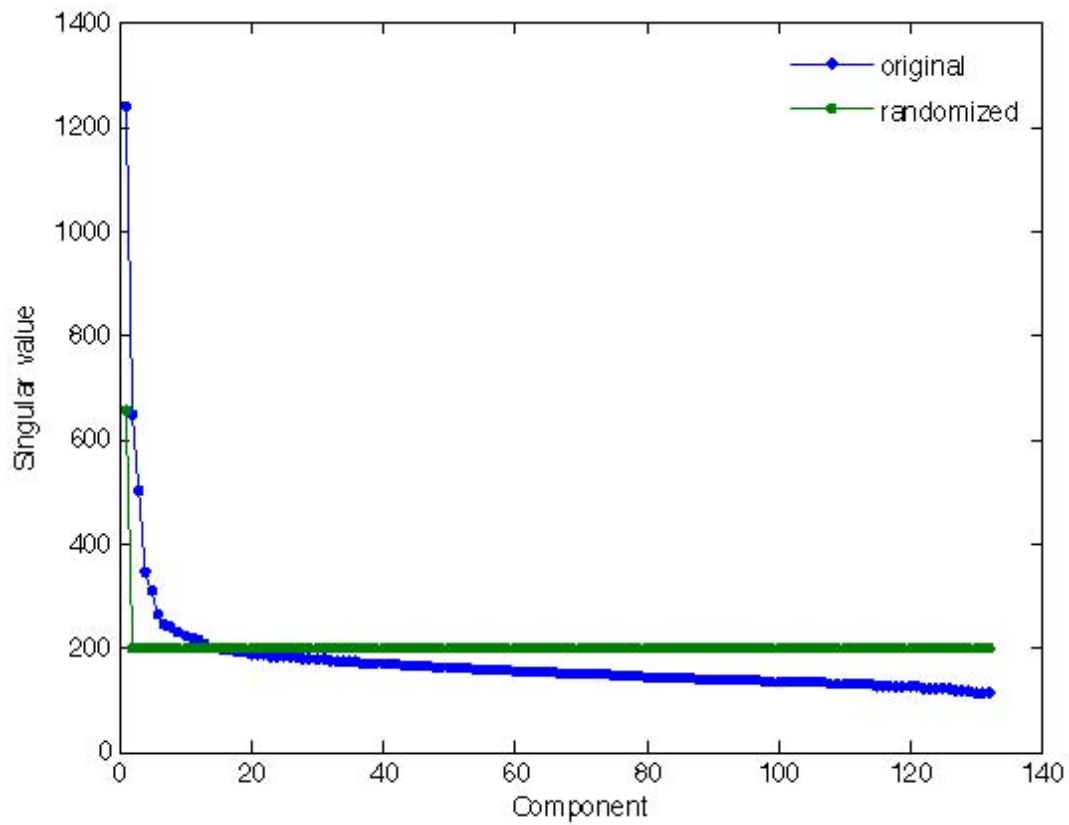

**Figure 3.** Singular values for 132 self-self hybridizations before (blue) and after (green) within-row-permutation. To find the number of major principal components, we randomly permuted ratio values for each probe and recomputed the singular values of the resulting matrix. The intersection between the original and permuted singular values occurs at component 14.

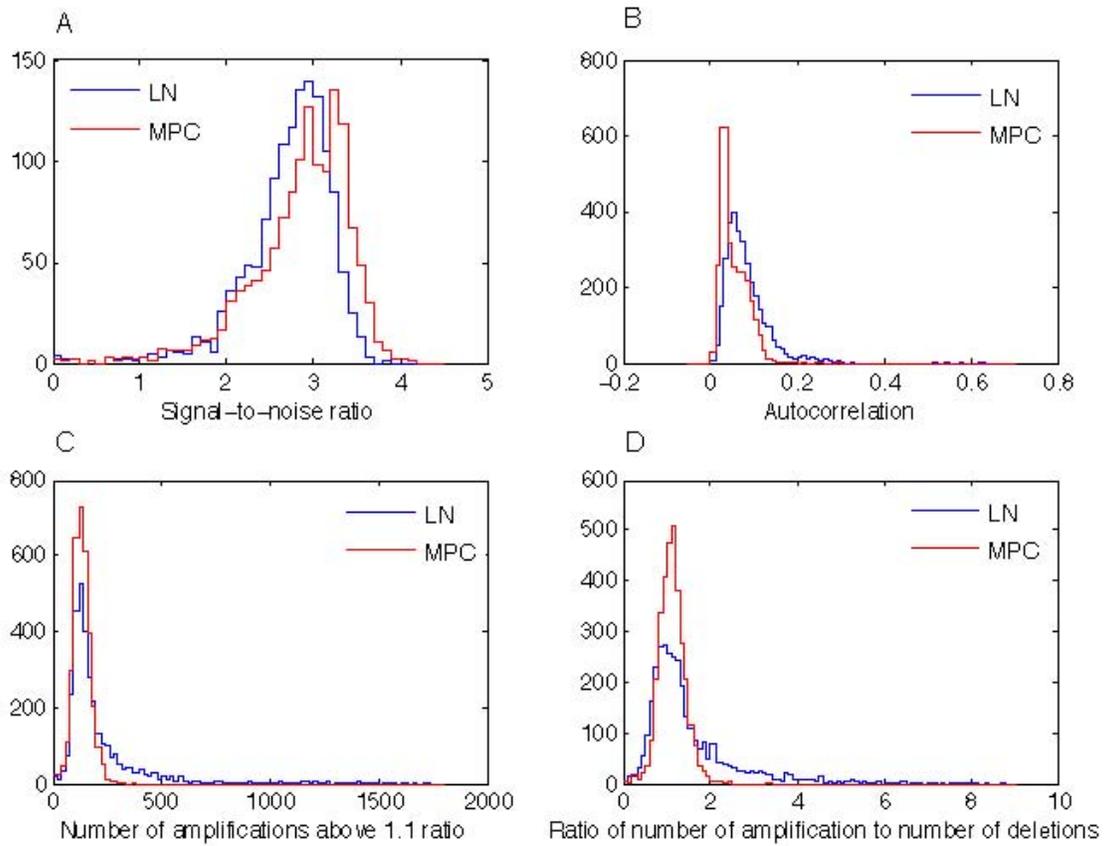

**Figure 4.** Measures of system noise in 3252 sample-reference hybridizations. Histograms of signal-to-noise ratio (A), autocorrelation (B), amplifications above 1.1 ratio threshold (C), and the ratio of number of amplifications to deletions with absolute ratio value above 1.1 threshold (D) are shown.

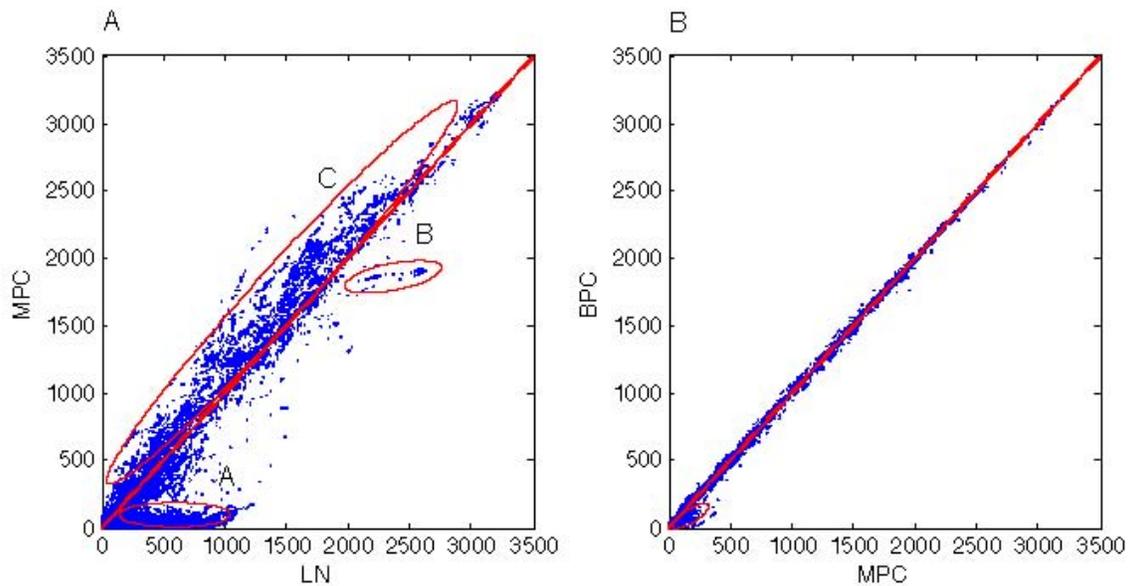

**Figure 5.** Comparison of normalization methods in sample-reference hybridizations. Data for probes on all autosomes, before and after normalizations, were segmented from 3252 hybridizations, median segmented ratio values assigned to each probe, and values above a 1.1 ratio threshold were counted. (A) Amplification count, with local normalization (X axis) vs. major principal component corrected data (Y axis). Circled region A represents the frequency of a large set of (low amplitude) segments detected before system normalization, which are drastically reduced after normalization; circled region B indicates the frequency of a subset of common copy number polymorphisms that are detected less frequently following MPC. Circled region C shows the common copy number polymorphisms that are detected more frequently following MPC. (B) Same as previous except comparing MPC (X axis) to BPC (Y axis). The circled region represents a small set of probes that are less frequently segmented, and for which the correction is improved.

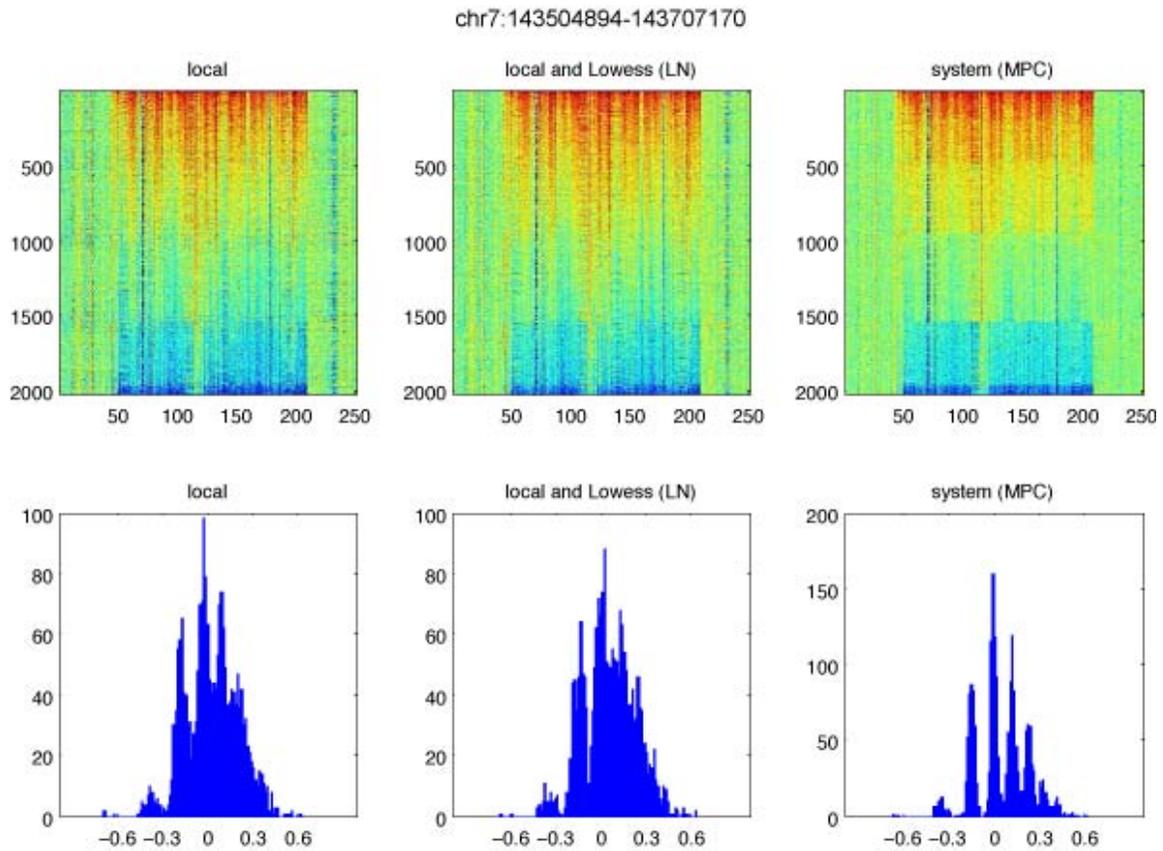

**Figure 6.** Discrete copy number states at a commonly polymorphic site after system normalization. The region displayed (chr7:143504894-143707170 in the hg18 build) consists of the CNP with forty non-polymorphic flanking probes on each side. The plots in the upper panels show the log ratio values of 2028 hybridizations (Y axis) for all probes in the extended region (X axis), in which the rows are sorted (in descending order) by their median probe ratio within the common interval. The lower plots are the corresponding histograms of the medians. Following local (left panels) and local/Lowess (LN) normalization (middle panels), varying copy number states are only moderately evident. System (MPC) normalization (rightmost top and bottom panels) resolves at least 6 distinct states at this locus.

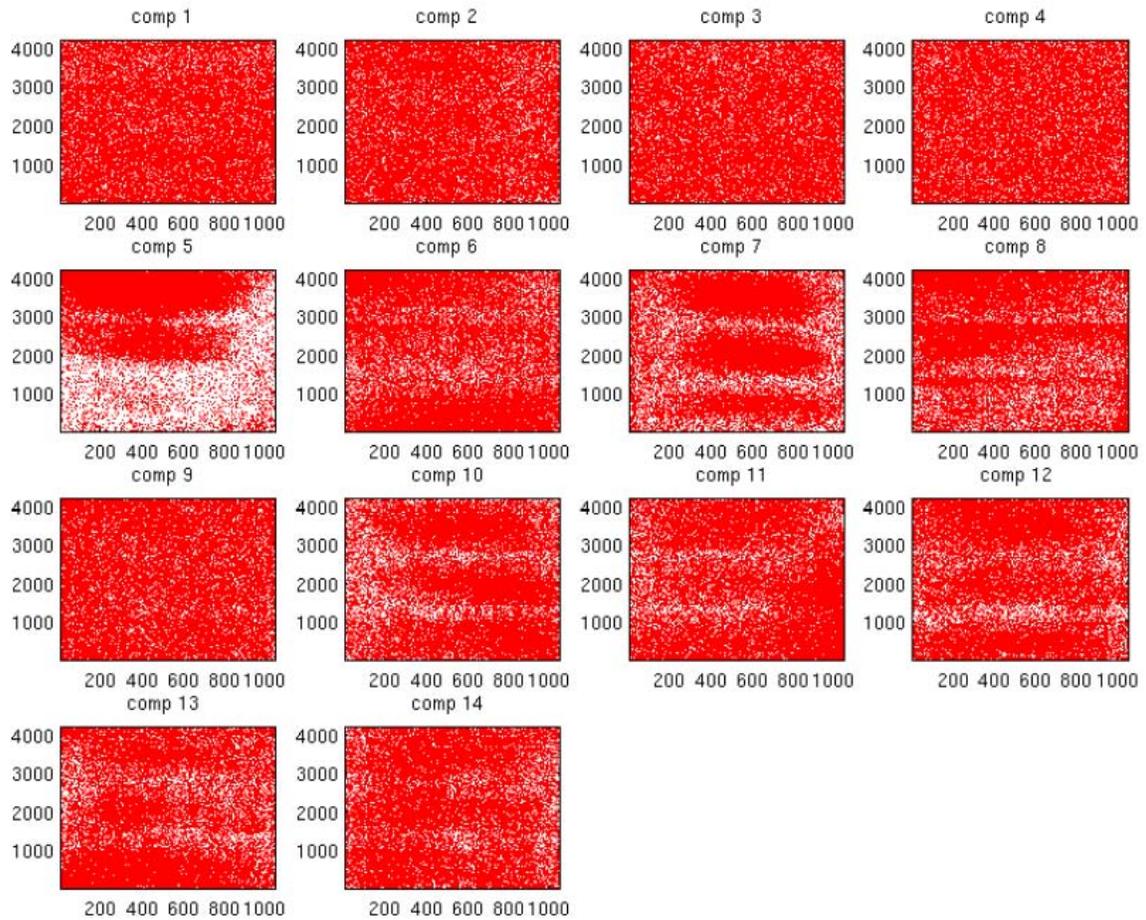

**Figure 7.** Patterns of spatial distribution of extreme probes on CGH microarrays. Array coordinates for probes with extreme loadings in each of 14 principal components (see text) are displayed as the X and Y axes in each of 14 plots. Spatial clustering for components 5–8 and 10–14 correspond to the three distinct hybridization and wash chambers for each array. Components 1-4 and 9 do not show any dependence on array coordinates.

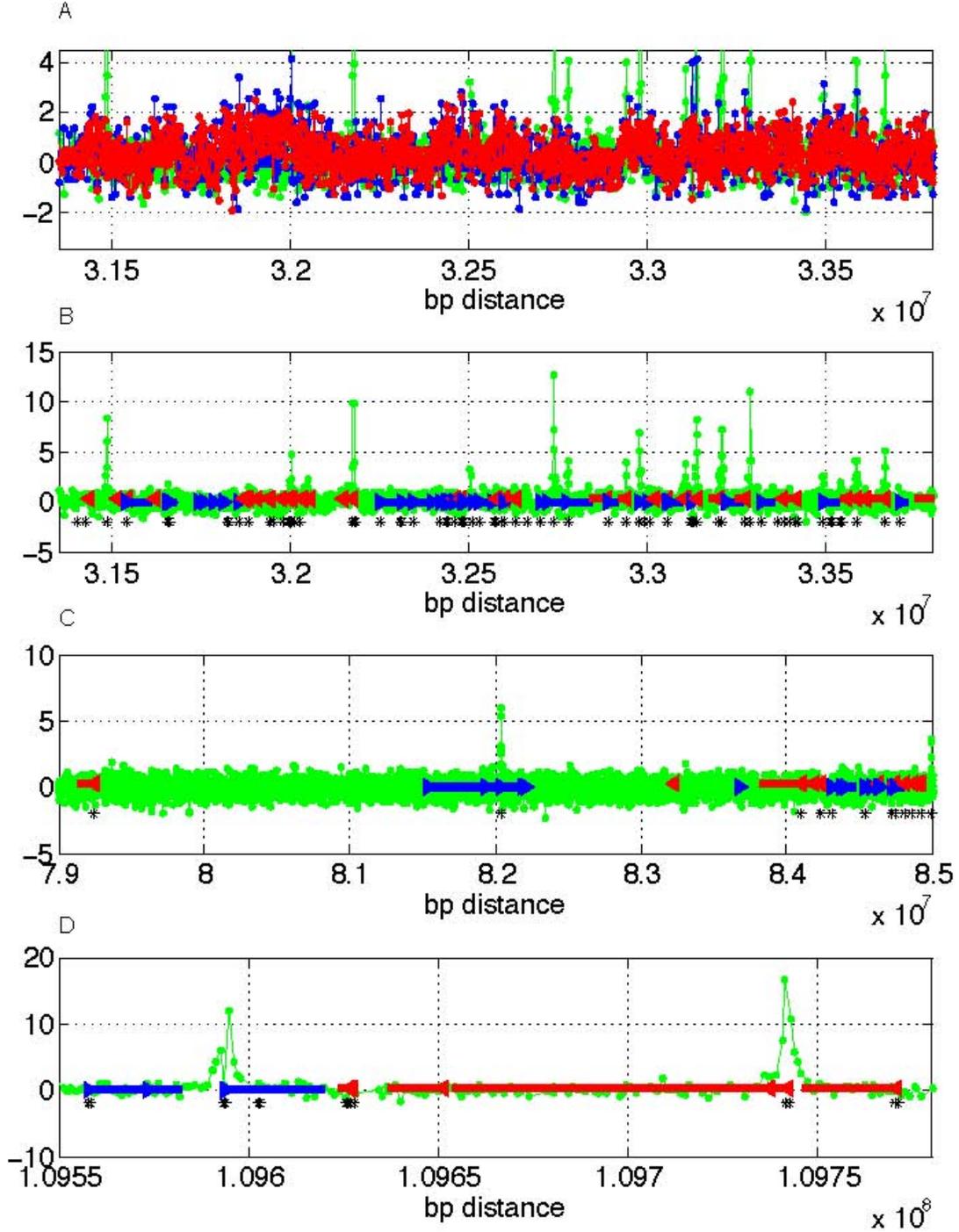

**Figure 8.** Genome order loadings from components 1 and 9, in relation to G/C content and gene transcription. The loadings of components 1 (red) and 9 (green) were examined in genome order from a representative gene-rich region of chromosome 1 (A). In the same panel, the (scaled) G/C composition for the corresponding probes (blue) is also shown. Red is on top of blue on top of green, illustrating the substantial coincidence of the plots for component 1 loadings and G/C content (and the lack thereof for component 9). Below, we illustrate the coincidence of peaks of component 9 loadings with respect to the genes in the same region (B). Green lines indicate loadings of component 9; blue and red represent forward- and reverse-strand genes, respectively; and the arrows indicate the direction of transcription and gene boundaries. Black asterisks show the genomic positions of CpG islands. The same relationship is also shown in different regions and at different scales (C,D). Probes with high loading from the ninth component form clusters about the 5´ ends of genes, especially those with nearby CpG islands. All information is derived from the hg18 build and UCSC Genome Browser (http://genome.ucsc.edu/). See text for further details.

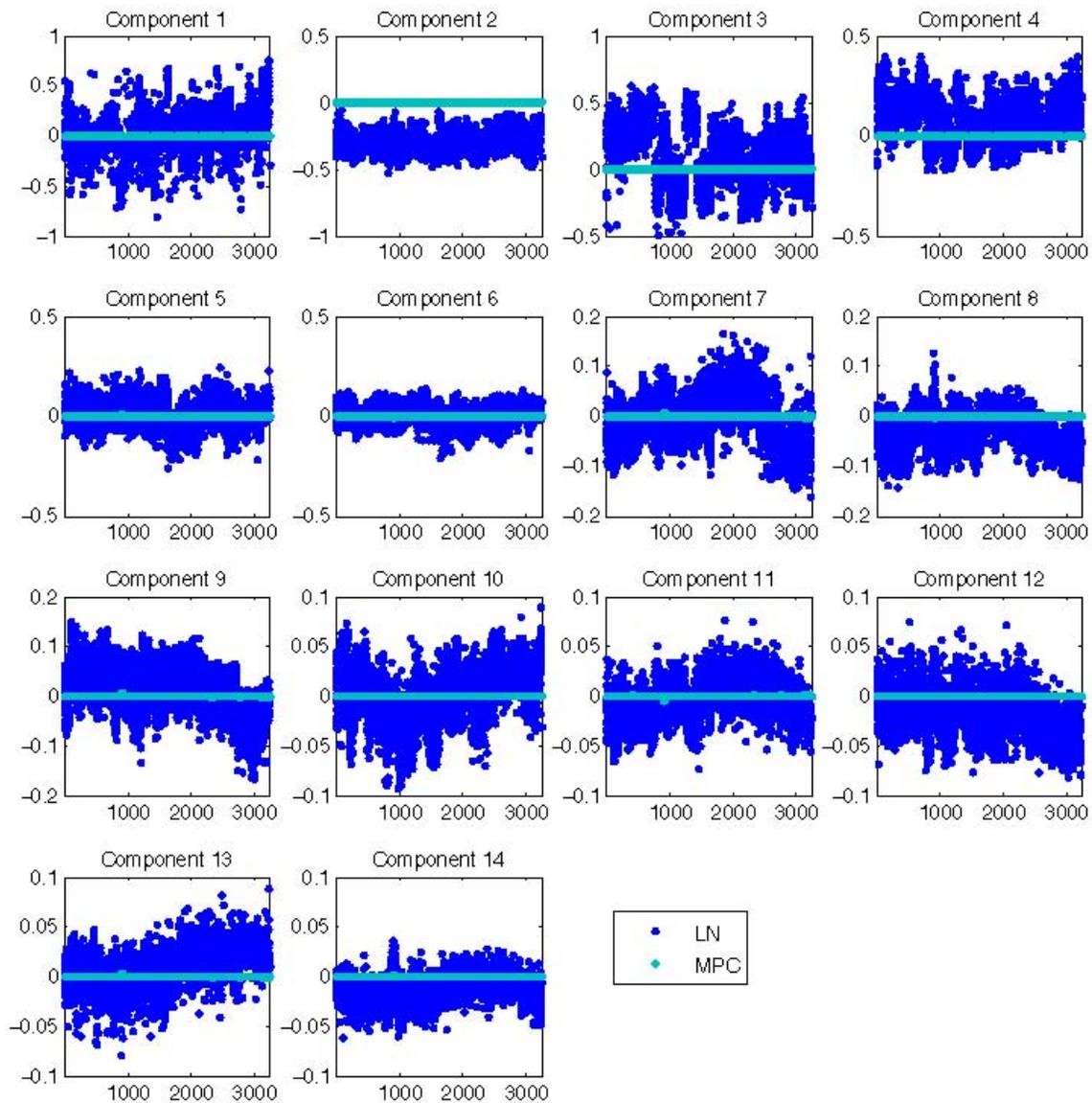

**Figure 9.** Correlation of major principal components with hybridizations over time. Each of 14 plots displays the correlation (Y axis) between the autosomal extreme loadings of the indicated principal component with the log ratios of the probes with those extreme loadings for each 3252 test-reference hybridizations, ordered by the 'queue index' (X axis). Dark blue represents local and Lowess normalization (LN) and light blue represents major principal component (MPC) normalization.

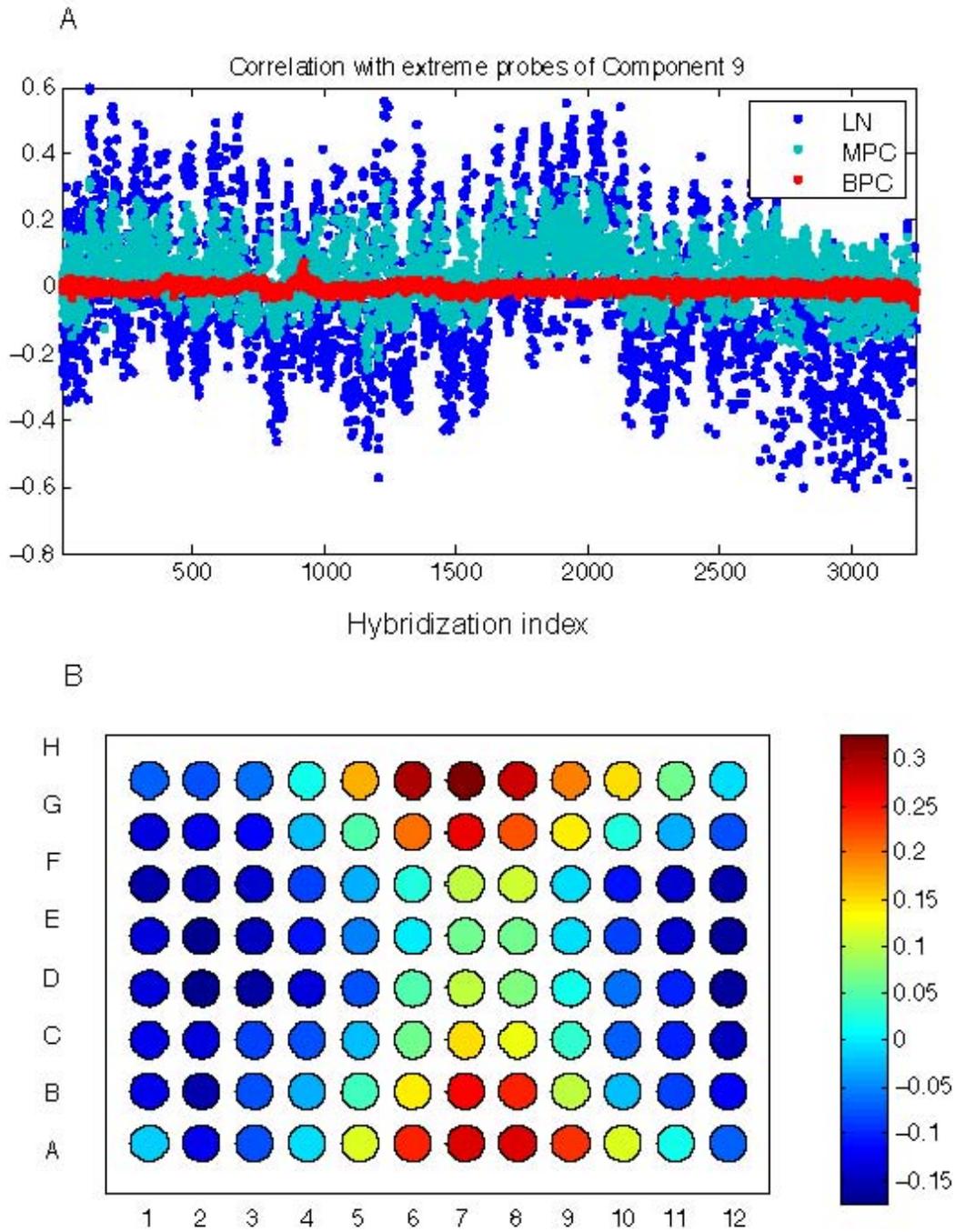

**Figure 10.** More peculiarities of the ninth component. (A) Periodicity of correlations with the component. As in Figure 9, we display the correlation between extreme loadings of component 9 and log ratios from 3252 hybridizations. Dark blue represents correlations with local and Lowess normalized (LN), light blue with major principal

component normalized (MPC), and red with batched principal component normalized (BPC) hybridizations. (B) Correlations with micro-well sample coordinates. The correlations computed and displayed in panel A with LN were adjusted for each 96-well plate to have a mean of zero and a standard deviation of 1. The adjusted values were then averaged over the same row and column coordinates from the 8x12 microwell plates in which the samples used for the hybridizations were stored and shipped. These values are then displayed in micro-well coordinates, with red for high positive and blue for high negative correlations.

**TABLES**

**Table 1. Characterization of principal components**

| Comp | SV | Skew | Kurtosis | AC | Autocorr | Total ND | GC |
|------|------|-------|----------|----|----------|----------|----|
| 1 | 1238.0 | -0.48 | 0.26 | - | 0.35 | 0.3183 | - |
| 2 | 648.6 | -0.14 | 0.54 | - | 0.05 | 0.1943 | - |
| 3 | 500.6 | -0.01 | 0.33 | - | 0.18 | 0.1368 | - |
| 4 | 346.1 | 0.23 | 0.62 | - | 0.09 | 0.1332 | - |
| 5 | 309.8 | 0.34 | 1.12 | + | 0.01 | 0.2222 | - |
| 6 | 263.7 | -0.02 | 0.42 | + | 0.03 | 0.0835 | - |
| 7 | 244.9 | 0.22 | 0.81 | + | 0.04 | 0.0964 | - |
| 8 | 237.4 | 0.13 | 0.93 | + | 0.03 | 0.0238 | - |
| 9 | 232.2 | 3.69 | 46.52 | - | 0.33 | 0.1149 | + |
| 10 | 223.2 | 0.07 | 0.53 | + | 0.02 | 0.0608 | - |
| 11 | 219.7 | 0.29 | 1.36 | + | 0.01 | 0.0929 | - |
| 12 | 214.9 | 0.10 | 0.70 | + | 0.02 | 0.1333 | - |
| 13 | 206.5 | 0.05 | 0.70 | + | 0.01 | 0.0318 | - |
| 14 | 199.6 | 0.01 | 0.55 | + | 0.01 | 0.0476 | - |

Column guide: Comp = component number; SV = singular value, diagonal values of matrix S in formula (2); Skew = skewness; Kurtosis = excessive kurtosis ; AC = array clustering observed in extreme loading values of component; Autocorr = autocorrelation, which is computed as correlation of the ratio vector shifted by one probe; Total ND = measure of nucleotide bias of probes with extreme loadings as described in Table 2, last column; GC = gene clustering (defined as the overlap of extreme probes with regions of gene transcription starts near CpG islands).

**Table 2. Compositional biases of principal components**

| Comp | A.neg | A.pos | C.neg | C.pos | G.neg | G.pos | T.neg | T.pos | Total Diff |
|---|---|---|---|---|---|---|---|---|---|
| 1 | **0.2981** | **0.2253** | **0.1998** | **0.2832** | **0.1829** | **0.2586** | **0.3192** | **0.2329** | **0.3183** |
| 2 | **0.3423** | **0.2451** | 0.2103 | 0.2185 | 0.2020 | 0.2038 | **0.2454** | **0.3326** | **0.1943** |
| 3 | **0.2416** | 0.2882 | **0.2635** | 0.2121 | **0.2226** | 0.2057 | **0.2723** | 0.2941 | **0.1368** |
| 4 | **0.2972** | **0.2306** | 0.2191 | **0.2575** | 0.2024 | **0.2281** | **0.2814** | 0.2838 | **0.1332** |
| 5 | **0.2624** | **0.3101** | **0.2385** | **0.1807** | **0.2230** | **0.1696** | **0.2761** | **0.3396** | **0.2222** |
| 6 | 0.2896 | 0.2727 | 0.2233 | 0.2058 | 0.2061 | 0.1987 | **0.2810** | **0.3228** | **0.0835** |
| 7 | 0.2719 | 0.2898 | 0.2266 | **0.1894** | 0.1960 | **0.1850** | 0.3055 | **0.3358** | **0.0964** |
| 8 | 0.2869 | 0.2844 | 0.2172 | 0.2106 | 0.1901 | **0.1872** | 0.3058 | **0.3177** | 0.0238 |
| 9 | **0.2475** | 0.2841 | 0.2199 | 0.2258 | 0.1937 | 0.2086 | **0.3390** | **0.2815** | **0.1149** |
| 10 | 0.2826 | 0.2869 | 0.2145 | **0.1982** | 0.1984 | **0.1842** | 0.3045 | **0.3307** | **0.0608** |
| 11 | **0.3011** | 0.2694 | **0.1831** | 0.2175 | **0.1841** | 0.1962 | **0.3316** | **0.3169** | **0.0929** |
| 12 | **0.2520** | **0.3186** | 0.2161 | **0.2000** | 0.1912 | **0.1886** | **0.3407** | 0.2928 | **0.1333** |
| 13 | 0.2866 | 0.2860 | 0.2127 | 0.2050 | **0.1823** | 0.1982 | **0.3184** | 0.3108 | 0.0318 |
| 14 | 0.2718 | 0.2943 | 0.2181 | 0.2038 | 0.1932 | 0.1946 | **0.3169** | 0.3073 | 0.0476 |
| Range in 1000 Random Simulations | [0.2643,0.2966] | | [0.2038,0.2324] | | [0.1888,0.2168] | | [0.2850,0.3155] | | [0,0.0587] |

For each of the 14 components, we computed the proportion of A, C, G or T in those probes with the bottom 1.5% negative ("neg") or top 1.5% positive ("pos") loadings. We computed the total difference ("Total Diff") in composition for probes with extreme positive and negative loadings, defined as the sum of the absolute values of the differences between x.pos and x.neg for each of the four bases. For each of 1000 simulations, a random subset of 1.5% of probes (32425 probes/simulation) was created and the range of the proportions of the four nucleotides was computed, creating the confidence intervals (p=10^-3) shown in the bottom row of cells. The bottom right cell contains the confidence interval (p=10^-3) of the total difference computed over 1000 pairs of random subsets of 32425 probes. Compositions outside the range (p < 10^-3) are in bold.

**Table 3. Probe cluster intervals overlapping with the 5´ ends of genes and/or CpG islands**

| Component | Cluster | Polarity | Gene & CpG | Gene Only | CpG Only |
|---|---|---|---|---|---|
| 1 | 714 | + | 0.10 | 0.08 | 0.22 |
| 1 | 30 | - | 0.00 | 0.00 | 0.00 |
| 2 | 143 | + | 0.55 | 0.02 | 0.11 |
| 2 | 59 | - | 0.05 | 0.05 | 0.19 |
| 3 | 111 | + | 0.00 | 0.04 | 0.01 |
| 3 | 399 | - | 0.17 | 0.06 | 0.35 |
| 4 | 484 | + | 0.12 | 0.11 | 0.20 |
| 4 | 22 | - | 0.18 | 0.05 | 0.14 |
| 5 | 34 | + | 0.00 | 0.00 | 0.00 |
| 5 | 9 | - | 0.11 | 0.00 | 0.22 |
| 6 | 36 | + | 0.44 | 0.03 | 0.08 |
| 6 | 20 | - | 0.10 | 0.00 | 0.10 |
| 7 | 117 | + | 0.50 | 0.01 | 0.15 |
| 7 | 22 | - | 0.00 | 0.05 | 0.00 |
| 8 | 83 | + | 0.59 | 0.05 | 0.14 |
| 8 | 34 | - | 0.03 | 0.00 | 0.03 |
| 9 | 3415 | + | 0.54 | 0.03 | 0.14 |
| 9 | 11 | - | 0.00 | 0.00 | 0.00 |
| 10 | 24 | + | 0.13 | 0.00 | 0.08 |
| 10 | 15 | - | 0.00 | 0.07 | 0.00 |
| 11 | 19 | + | 0.16 | 0.00 | 0.00 |
| 11 | 28 | - | 0.00 | 0.00 | 0.00 |
| 12 | 19 | + | 0.00 | 0.00 | 0.00 |
| 12 | 55 | - | 0.51 | 0.04 | 0.04 |
| 13 | 30 | + | 0.03 | 0.03 | 0.00 |
| 13 | 8 | - | 0.00 | 0.00 | 0.00 |
| 14 | 25 | + | 0.00 | 0.04 | 0.00 |
| 14 | 21 | - | 0.19 | 0.05 | 0.10 |

The column "Cluster" shows the total number of probe cluster intervals, defined as a maximally contiguous set of at least three probes from the top (+) and bottom (-) extreme 1.5% of loadings ("Polarity"), from each major component. The last three columns show the proportion of probe cluster intervals from the top and bottom extreme loadings for each principal component that overlap both 5´ ends of genes and CpG islands ("Gene & CpG"), 5´ end of genes only ("Gene Only"), and CpG islands only ("CpG Only"). While many components have clusters similar to component 9 in the proportion distributing to the 5'ends of genes and CpG islands, none have these clusters in such abundance.